\documentclass[%
reprint,
superscriptaddress,
 amsmath,amssymb,
 pra,
aps
]{revtex4-1}

\usepackage[utf8]{inputenc}
\usepackage[hidelinks]{hyperref}
\usepackage[english]{babel}
\usepackage{libertine}
\usepackage{libertinust1math}
\usepackage{graphicx}
\usepackage{dcolumn}
\usepackage{bm}
\usepackage{upgreek}

\begin{document}
\title{On-Chip Integration of Single Solid-State Quantum Emitters with a \texorpdfstring{SiO$_2$}{} Photonic Platform}


\author{Florian B\"ohm}
\email[Author e-mail address: ]{Florian.Boehm@physik.hu-berlin.de}
\affiliation{Institut f\"ur Physik, Humboldt-Universit\"at zu Berlin, Newtonstr. 15, D-12489 Berlin, Germany}
\affiliation{IRIS Adlershof, Humboldt-Universit\"at zu Berlin, Zum Großen Windkanal 6, D-12489 Berlin, Germany}

\author{Niko Nikolay}
\affiliation{Institut f\"ur Physik, Humboldt-Universit\"at zu Berlin, Newtonstr. 15, D-12489 Berlin, Germany}
\affiliation{IRIS Adlershof, Humboldt-Universit\"at zu Berlin, Zum Großen Windkanal 6, D-12489 Berlin, Germany}
 
\author{Christoph Pyrlik}
\affiliation{Ferdinand-Braun-Institut, Leibniz-Institut für H\"ochstfrequenztechnik, Gustav-Kirchhoff-Str. 4, D-12489 Berlin, Germany}

\author{Jan Schlegel} 
\affiliation{Ferdinand-Braun-Institut, Leibniz-Institut für H\"ochstfrequenztechnik, Gustav-Kirchhoff-Str. 4, D-12489 Berlin, Germany}

\author{Andreas Thies} 
\affiliation{Ferdinand-Braun-Institut, Leibniz-Institut für H\"ochstfrequenztechnik, Gustav-Kirchhoff-Str. 4, D-12489 Berlin, Germany}

\author{Andreas Wicht} 
\affiliation{Ferdinand-Braun-Institut, Leibniz-Institut für H\"ochstfrequenztechnik, Gustav-Kirchhoff-Str. 4, D-12489 Berlin, Germany}

\author{G\"unther Tr\"ankle} 
\affiliation{Ferdinand-Braun-Institut, Leibniz-Institut für H\"ochstfrequenztechnik, Gustav-Kirchhoff-Str. 4, D-12489 Berlin, Germany}

\author{Oliver Benson}
\affiliation{Institut f\"ur Physik, Humboldt-Universit\"at zu Berlin, Newtonstr. 15, D-12489 Berlin, Germany}
\affiliation{IRIS Adlershof, Humboldt-Universit\"at zu Berlin, Zum Großen Windkanal 6, D-12489 Berlin, Germany}

\date{\today}

\begin{abstract}

One important building block for future integrated nanophotonic devices is the scalable on-chip interfacing of single photon emitters and quantum memories with single optical modes. Here we present the deterministic integration of a single solid-state qubit, the nitrogen-vacancy (NV) center, with a photonic platform consisting exclusively of SiO$_2$ grown thermally on a Si substrate. The platform stands out by its ultra-low fluorescence and the ability to produce various passive structures such as high-Q microresonators and mode-size converters. By numerical analysis an optimal structure for the efficient coupling of a dipole emitter to the guided mode could be determined. Experimentally, the integration of a preselected NV emitter was performed with an atomic force microscope and the on-chip excitation of the quantum emitter as well as the coupling of single photons to the guided mode of the integrated structure could be demonstrated. Our approach shows the potential of this platform as a robust nanoscale interface of on-chip photonic structures with solid-state qubits.
\end{abstract}

\maketitle

\section{\label{sec:intro}Introduction}
Single solid-state quantum emitters such as color centers in diamond are promising potential building blocks for future quantum information processing architectures and integrated nanophotonic devices \cite{Aharonovich2014}. 
The most prominent representative is the nitrogen-vacancy (NV) color center which exhibits coherent optical transitions and long-lived nuclear and electron spins, making it a promising solid-state qubit \cite{Gruber1997,Togan2010,Maurer2012} and single photon source \cite{Kurtsiefer2000}, but the ongoing search for new solid-state quantum emitters recently revealed a variety of new emitters, e.g. new optical vacancy-impurity defects in diamond \cite{Thiering2018,Luhmann2018} and hexagonal boron nitride \cite{Tran2016}.

The most fundamental way of interaction with single emitters is free-space optics, but in order to build scalable quantum architectures, deterministic and efficient on-chip integration of one or more quantum emitters in combination with passive optical structures such as cavities and couplers is required \cite{Kimble2008,Benson2011}.
One fundamental requirement is the efficient collection and routing of single photons. The commonly used platform silicon cannot be used in most cases since its opaque up to about $1.1\,\mathrm{\upmu m}$ and a large part of the interesting solid-state emitters emit at visible wavelengths
Therefore, in recent years great efforts have been made to develop new platforms for integration of quantum emitters and to demonstrate the assembly of hybrid nanophotonic systems, for example, based on tapered optical fibers \cite{Schroder2012,Liebermeister2014,Fujiwara2016,Patel2016,Schell2017}, in-situ direct laser written \cite{Shi2016} or electron beam lithographed {\cite{Schnauber2018}}, diamond \cite{Faraon2011,Burek2017,Bhaskar2017}, other dielectric \cite{Gschrey2013,Zadeh2016,Gould2016,Kim2017a,Tonndorf2017,Mouradian2015,Davanco2017} or plasmonic structures \cite{Kewes2016,Siampour2017,Siampour2018}.

Each of these platforms poses its own challenges and depending on the selected emitter and specific application, some are better suited than others. One common problem is unwanted background fluorescence from the passive optical structure, as some emitters need to be pumped strongly with green or even blue light. Another crucial aspect is low absorption in the material to minimize losses and enable the fabrication of high-Q resonators.
A very well-suited material platform is silica (SiO$_2$) which is thermally grown on silicon (Si), as it offers a broad transparency window from the UV to mid-IR and therefore low fluorescence and losses. However, the index of refraction of SiO$_2$ is rather low ($\sim 1.5$), so in order to allow for a propagating mode within the SiO$_2$, the underlying silicon must to be removed.

In this work we use a free-standing, monolithic SiO$_2$ photonics platform, where rib-waveguides allow efficient guiding of visible light in thermally grown and undoped SiO$_2$ similar to the system introduced by Chen \textit{et al.} \cite{Chen2012}. So far, we have been able to show that this system has a very low background fluorescence, which is well suited for the integration of single photon emitters, and in addition high-Q micro-resonators can be realized \cite{Pyrlik2019}. By the numerical analysis of the waveguide structures we present a design that allows an optimal coupling of the guided mode in the waveguide to an external quantum emitter. We also present a 2D tapered section of the waveguide as a mode-size converter to improve the mode overlap with lensed single-mode fibers. After characterizing the fabricated device, we show the functionalization of this nanophotonic platform with a preselected single NV color center hosted within a nanometer-sized diamond \cite{Schell2011}, and by detecting the single-photon emission from the integrated waveguide, we were able to demonstrate the coupling between the guided fundamental mode of the SiO$_2$ waveguide and the single quantum emitter.

The integrated platform presented here has the advantage of the ultra-low intrinsic fluorescence when compared to other material such as Si$_3$N$_4$ or doped optical fibers. Its wide transparency window ($\sim 0.2\,\mathrm{\upmu m}$ - $3.0\,\upmu\mathrm{m}$) is ideal to integrate other solid-state or condensed-phase emitters, e.g. in a hybrid integrated platform. Finally, fabrication is simple and other passive structures such as high-Q microring resonators \cite{Pyrlik2019} or detectors \cite{Ferrari2018} can be integrated.

\section{System and methods}
\subsection{\label{sec:wgdesign}Waveguide design and functionalization}
\begin{figure*}[htbp]
  \centering
  \includegraphics[width=0.8\textwidth]{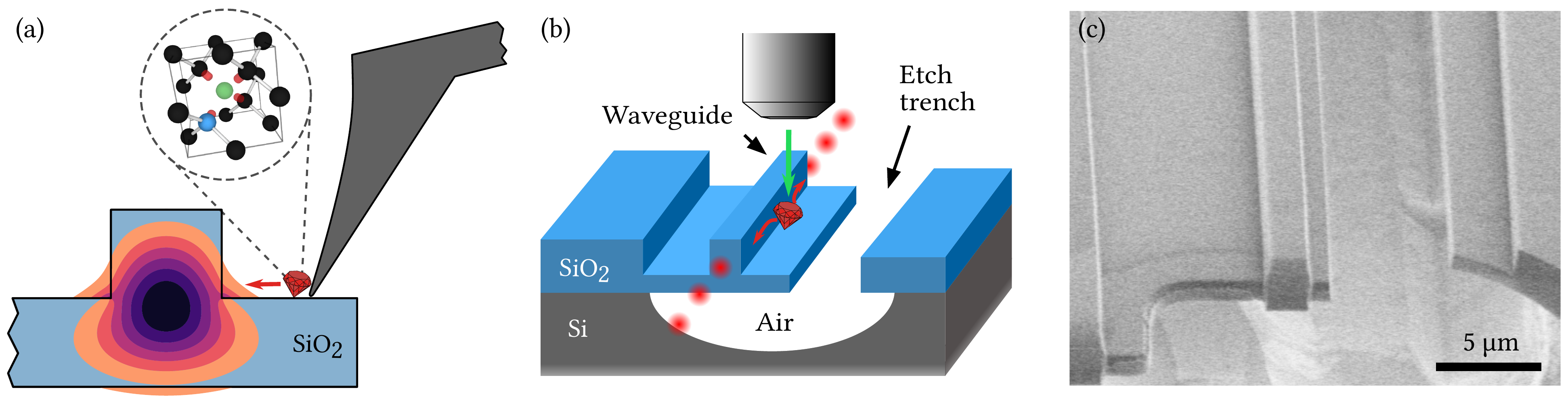}
\caption{\textit{Waveguide design and functionalization.}
(a) Illustration of the SiO$_2$ waveguide structure and the field profile ($|E|^2$) of the guided TM fundamental optical mode at $700\,{nm}$. Also the deterministic positioning process of the diamond-nanocrystal containing a single NV center (the NV crystal structure is shown in the inset) into the inner edge of the integrated SiO$_2$ rib waveguide with an atomic force microscope (AFM) tip, is shown. 
(b) Schematic of the assembled device with the single quantum emitter at its desired location within the inner edge of the waveguide, evanescently coupled to the guided mode. The schematic also points out the underetched freestanding rib waveguide, which allows mode guiding in pure SiO$_2$ and one possible excitation/detection scheme.
(c) Scanning electron microscope image of the fabricated structure. The waveguide is recessed to prevent damage to the facet by dicing during fabrication of the device and also helps prevent damaging the freestanding rib waveguide during the experiments.
}
\label{fig:fig1}
\end{figure*}

To minimize background fluorescence, the integrated photonic platform was designed to guide the optical mode exclusively within undoped, thermally grown SiO$_2$ which exhibits ultra-low intrinsic fluorescence, even when strongly pumped with a $532\,\mathrm{nm}$ laser \cite{Pyrlik2019}. 
Single NV centers on fabricated devices typically show signal-to-noise ratios of  $\approx 45$ (with subtracted dark count rates).
To enable this air-clad waveguide, a supporting membrane is required, resulting in the rib-waveguide structure schematically shown in \hyperref[fig:fig1]{\autoref{fig:fig1}(a)}. These waveguides usually support at least two fundamental modes with orthogonal electric fields, one almost purely transversal electric (quasi-TE, hereafter referred to as TE) and one mostly transversal magnetic (quasi-TM, hereafter referred to as TM). The cross section shows the field profile ($|E|^2$) of the guided TM fundamental optical mode, which visualizes the evanescent tail of the mode within the inner edge of the rib waveguide. Here we also sketch the positioning of a nanodiamond of some ten nanometers in size hosting a single NV center to the designated position within the inner edge of the rib waveguide. As it is evident from the field profile, this is the best accessible spot to place an optical emitter in the evanescent field of the guided mode. In order to have a significant part of the evanescent field accessible the dimensions of the waveguide should be in the range or smaller than the wavelength of the guided light ($\lambda\approx 700\,\mathrm{nm}$). 

A nanoparticle hosting a single NV center can be pre-characterized and transferred to the integrated structure by pick-and-place manipulation using a commercial atomic force microscope (AFM; JPK Instruments) as elucidated in \cite{Schell2011}. Since the Silicon substrate is opaque to visible light, the positioning and verification of placement of the NV center could not be done in-situ, but rather the diamond was placed some micrometers next to the waveguide. After optically verifying the successful placement of the nanodiamond hosting a single NV center by performing a confocal scan, the nanoparticle is pushed to the inner edge of the rib waveguide in a subsequent step, using a specially shaped tip (AdvancedTEC\texttrademark  NC, NANOSENSORS\texttrademark). Similarly to Ref. \cite{Schell2011}, the yield of a successful pick-and-place process with subsequent manipulation towards the desired position was about $1/3$.

In \hyperref[fig:fig1]{\autoref{fig:fig1}(b)} the assembled device is illustrated in more detail, also showing the silicon substrate that is removed underneath the waveguide during the fabrication process, which will be presented elsewhere in detail \cite{Thies}. In the figure we also present a possible experimental configuration in which the deposited nanodiamond emitter is excited by free-space pumping through a microscope objective and the emitted single photons are evanescently coupled to the guided modes of the waveguide, which will be discussed in more detail in the next section.

\hyperref[fig:fig1]{\autoref{fig:fig1}(c)} shows a scanning electron microscope image of the integrated free-standing rib waveguide structure. The waveguide is slightly recessed to prevent damage to the facet during the fabrication process and also proves to be advantageous during the experiments, as it provides some protection for the waveguide's facet.

\subsection{\label{sec:guidedmodes}Guided modes and coupling efficiency}
\begin{figure*}
  \centering
  \includegraphics[width=0.8\textwidth]{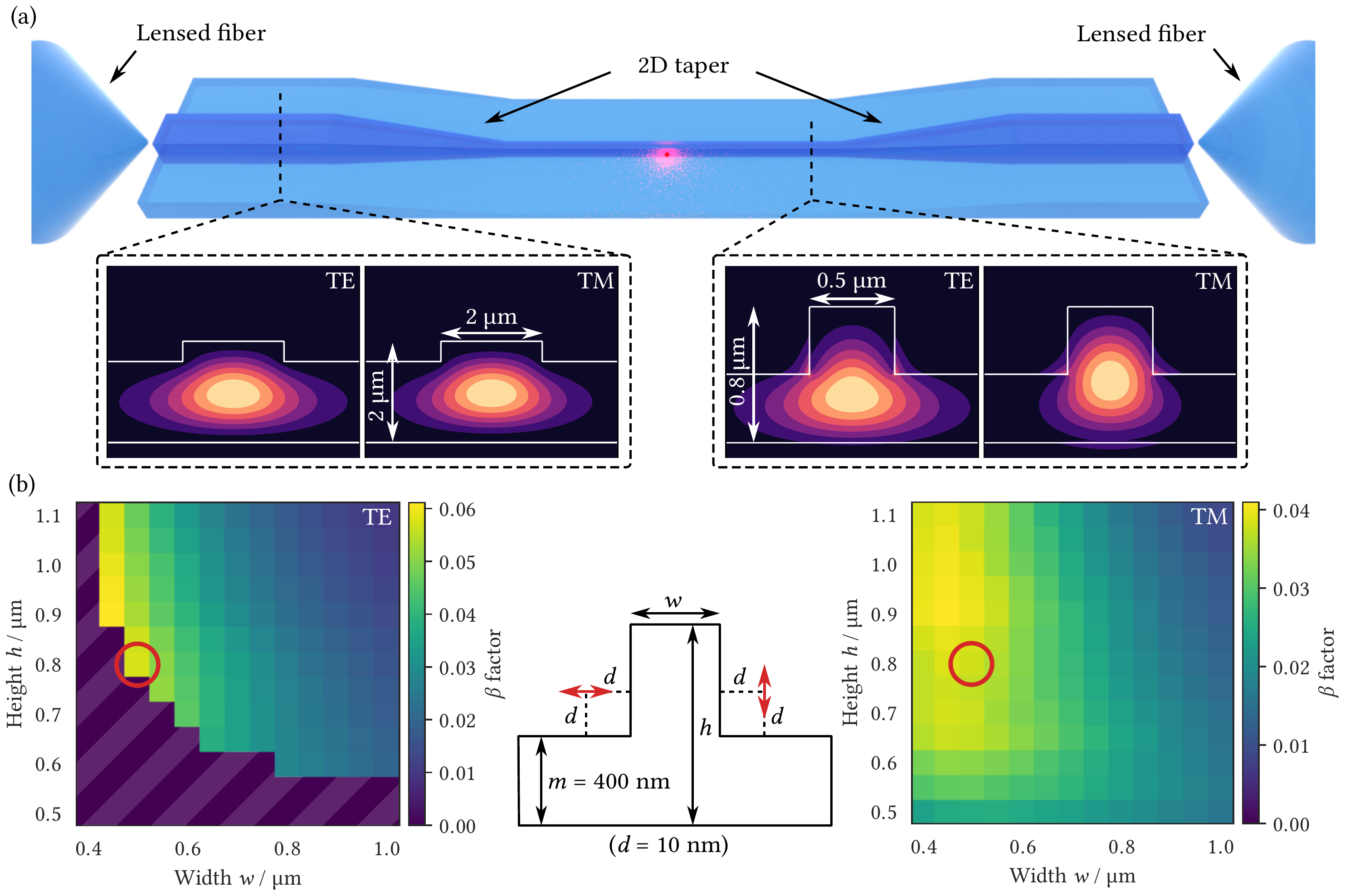}
\caption{\textit{Guided modes and coupling efficiency.}
(a) Top: Schematic of the waveguide-fiber coupling and lateral and vertical (2D) tapered waveguide section to improve the coupling between the waveguide and lensed single-mode fibers. Bottom: For the two waveguide cross-sections the field profile ($|E|^2$) of the fundamental TE and TM mode at a wavelength of $\lambda=700\,\mathrm{nm}$ obtained from eigenmode simulations are represented.  (b) Calculation of the $\beta$ factor. Center: Simulation design, the dipole emitter is  located $d=10\,\mathrm{nm}$ from either side of the inner edge of the SiO$_2$ rib waveguide with a membrane thickness $m = 400\,\mathrm{nm}$. The dipole axis of the emitter is aligned either horizontally or vertically to the waveguide as indicated by the red arrows. Left (right): $\beta$ factor of the horizontal (vertical) dipole with the TE (TM) mode. The shaded area indicates where TE modes become lossy for smaller dimensions of the waveguide. The final waveguide dimensions chosen to be fabricated as a device ($w=500\,\mathrm{nm}$, $h=800\,\mathrm{nm}$) are marked with red circles.
These parameters were chosen as a trade-off between the optimal coupling efficiency to the guided mode and fabrication simplicity.
}
\label{fig:fig2}
\end{figure*}

To conveniently couple the light in and out of the waveguide chip, we chose silica core single mode lensed fibers (S630-HP) with a spot diameter of
$(2.0\pm0.5)\,\mathrm{\upmu m}$
(OZ Optics). Due to the mismatch between the fiber's spot size and the dimensions of the mode in the emitter coupling region of the waveguide (sub-micron) we designed a mode-size converter. The inverse tapers usually employed in strip waveguides for mode conversion \cite{Almeida2003} are not feasible for our rib waveguide structure due to the requirement of the supporting membrane, therefore a two-dimensional (2D) section of the waveguide is tapered both laterally and vertically, as illustrated in \hyperref[fig:fig2]{\autoref{fig:fig2}(a)}. This tapered section acts as an adiabatic mode size converter and increases the on- and off-chip coupling efficiency with the single mode fibers. The fundamental TE and TM eigenmode profiles at a wavelength of $\lambda=700\,\mathrm{nm}$ before and after the 2D tapered section are shown in the lower part of \hyperref[fig:fig2]{\autoref{fig:fig2}(a)}. The eigenmode solutions are obtained from a frequency-domain guided-mode solver (JCMwave).
When calculating the overlap of the guided mode in the waveguide and the spot diameter of the lensed fiber, we can find a power coupling of $(20^{+11}_{-6})\,\%$ and  $(15^{+9}_{-5})\,\%$ without the tapered waveguide section for TE and TM polarization respectively. When the waveguide size is increased by a 2D taper, the power coupling to the fundamental mode is increased to $(58\pm 10)\,\%$ and $(57\pm 11)\,\%$ for TE and TM, respectively.

To experimentally estimate the on-chip coupling efficiency, we used a laser with an emission wavelength of $637\,\mathrm{nm}$, which was coupled into the waveguide from one side with the lensed fiber and collected the guided light on the other side of the waveguide chip with an objective lens. The lensed fiber was mounted with a V-groove fiber holder, coarse and fine positioning was performed with a XYZ linear translation stage (Thorlabs) and a XYZ piezo positioning stage (PiezoJena) respectively. After careful optimization of the fibers position with respect to the waveguide, the transmitted light is recorded on a photodetector and its power is compared with the laser power measured directly behind the lensed fiber. A fiber beam splitter is used to generate a power reference signal in order to take into account fluctuations in laser power within the fiber. In addition, transmitted light polarization can be adjusted with an inline fiber polarization controller to be mainly TE or TM polarized. The ratio of transmitted light to incident light can then be used as lower bound for on-chip coupling efficiency, assuming zero losses within the waveguide and tapered sections.

Each waveguide chip was fabricated containing a variety of waveguides with different dimensions. A total of 14 waveguides with 2D tapers and dimensions comparable to those in \hyperref[fig:fig2]{\autoref{fig:fig2}(a)} were tested, and apart from some outliers likely due to defects in these waveguides, the average total measured transmission was $(32.5 \pm 3.4)\,\%$ for both incident polarization angles. The deviation of the calculated power coupling value from the experimentally obtained values for transmission is most likely due to a combination of losses, the imperfect orientation of the fiber with respect to the waveguide and propagation losses within the waveguide. We would like to emphasize that we did not observe significant differences comparing transmission measurements done with tapered and untapered waveguides, hence we assume that tapering of the waveguides does not lead to notable additional losses.

The interaction efficiency of the emitter with the guided modes in the waveguide can be described by the $\beta$ factor, defined by the fraction of the total emitted energy which is coupled to the guided mode $\beta=\frac{\varGamma_\mathrm{wg}}{\varGamma_\mathrm{tot}}$, where $\varGamma_\mathrm{wg}$ denotes the emitters decay rate into the guided modes of the waveguide and $\varGamma_\mathrm{tot}$ the emitter’s total decay rate.
The $\beta$ factor can be obtained from performing a full 3D simulation of the waveguide structure with a radiating dipole emitter where the output into the guided mode is monitored. 
To facilitate the problem, we exploit the possibility of calculating the coupling Purcell factor $P=\frac{\varGamma_\mathrm{wg}}{\varGamma_\mathrm{0}}$ for a point-like dipolar emitter from a simple 2D simulation of the guided modes as described by Barthez \textit{et al.} \cite{Barthes2011}:
\begin{equation}
P=\frac{\varGamma_\mathrm{wg}}{\varGamma_\mathrm{0}}=\frac{3\pi c E_\mathrm{u}(r)\left[E_\mathrm{u}(r)\right]^*}{k_0^2\int_{A \infty}(\vec{E}\times\vec{H}^*)\cdot \hat{z}dA}
\end{equation}
where $\varGamma_\mathrm{0}$ denotes the undisturbed decay rate of the emitter (in vacuum), $\hat{z}$ is a normalized vector pointing in direction of propagation along the waveguide which is normal to the surface $dA$, $k_0$ is the absolute value of the photon momentum in air and $E_\mathrm{u}$ denotes the electric field components parallel to the dipole orientation of an emitter. 
The layout for the simulation can be found in the center of \hyperref[fig:fig2]{\autoref{fig:fig2}(b)} where the dipole emitter was placed either vertical or horizontal in respect to the waveguide structure $d=10\,\mathrm{nm}$ from each side of the inner edge.


In order to calculate $\beta$ from $P$ a 3D calculation of the same structure is carried out, from which the total emitted power of the radiating dipolar point source sitting next to the waveguide is obtained as well as the total emitted power in vacuum. 
By combining the results of the 2D simulation which gives us $P$ and the 3D simulation from which we obtain $\varGamma_\mathrm{0} / \varGamma_\mathrm{tot}$ we can then calculate $\beta$ as:
\begin{equation}
\beta=\frac{P\cdot\varGamma_\mathrm{0}}{\varGamma_\mathrm{tot}}
\end{equation}

The dimensions where $\beta$ is optimal are found by a parameter scan of the height $h$ and width $w$ of the waveguide. In the scan the thickness of the membrane was set constant to a value of $m = 400\,\mathrm{nm}$ where a robust fabrication and handling of the integrated structures was still ensured and the wavelength was set to $\lambda = 700\,\mathrm{nm}$ as this is approximately the peak of the NV centers emission. 
The nanometer-sized diamonds chosen for integration typically have a diameter of $\sim 20\,\mathrm{nm}$ and the NV center was assumed to be located at the center of the nanocrystal.
The $\beta$ factor for the fundamental TE (TM) guided mode coupling to a horizontal (vertical) dipole emitter $d=10\,\mathrm{nm}$ from the inner edges can be found in \hyperref[fig:fig2]{\autoref{fig:fig2}(b)} left (right). As TE modes become no longer strongly guided and therefore very lossy for smaller dimensions of the waveguide $\beta$ was set to zero here and the area is marked shaded.

The final device was designed to support both the fundamental TE and TM mode due to the random orientation of the NV centers dipole axis, so we chose a width $w=500\,\mathrm{nm}$, height $h=800\,\mathrm{nm}$ and membrane thicknesses $m = 400\,\mathrm{nm}$, marked with red circles in \hyperref[fig:fig2]{\autoref{fig:fig2}(b)}. The simulated $\beta$ factor for a device with these dimensions is $\beta_\mathrm{TE} =5.7\,\%$ for a horizontally oriented dipole coupling to the TE mode and $\beta_\mathrm{TM} = 3.8\,\%$ for a vertical oriented dipole coupling to the TM mode in the waveguide, with corresponding values of $P_\mathrm{TE} =0.14 $ and $P_\mathrm{TM} = 0.17$ for the coupling Purcell factor.

As shown in \hyperref[fig:fig2]{\autoref{fig:fig2}(b)}, a device with a higher aspect ratio could only marginally improve coupling efficiency, so the fabrication of such a device was not pursued. On the other hand, it is possible to significantly increase the coupling efficiency by reducing the membrane height, which improves access to the guided mode but decreases mechanical stability. A detailed discussion on increasing coupling efficiency can be found in \autoref{sec:increasing}.

\section{\label{sec:results}Results of the assembled device}
\begin{figure*}
  \centering
  \includegraphics[width=0.8\textwidth]{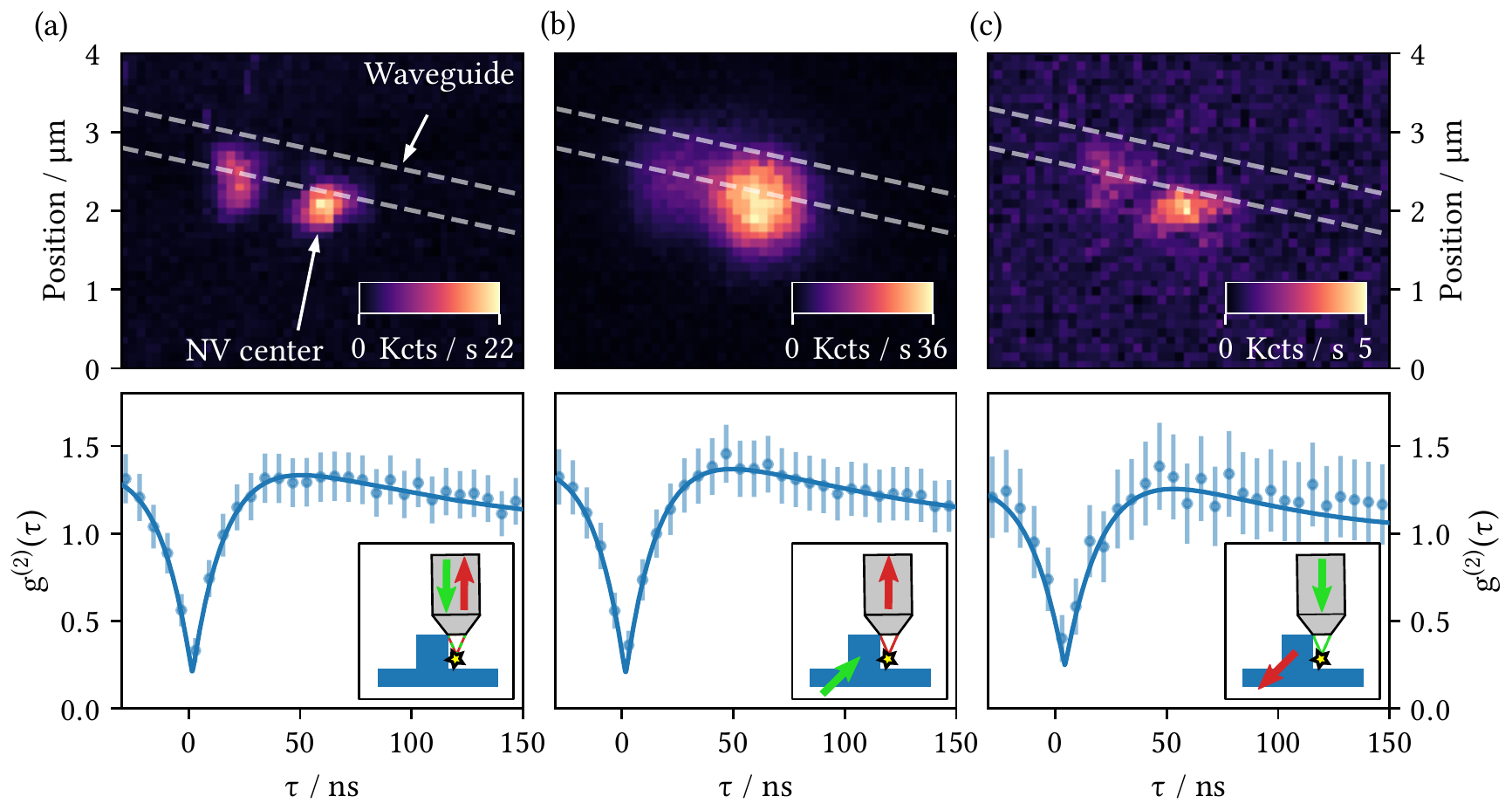}
\caption{
\textit{Results of the assembled device.}
Confocal fluorescence scans (top) and second-order autocorrelations (bottom) of the fluorescence detected from a single NV center coupled to a SiO$_2$ waveguide (approximate position indicated by the dashed lines) for the three different experimental configurations represented in the insets, i.e., (a): excitation and detection via the microscope objective, (b): excitation via the waveguide and detection via the microscope objective, (c): excitation via the objective and detection via the two output ports of the waveguide. For better visibility, each point in the autocorrelation measurement represents 25 binned measurement points and the error bars correspond to the square root of the coincidence events within each time bin. The solid line is the fit to a three-level model (see main text). The NV center is excited with a green ($532\,\mathrm{nm}$) laser source and the detection window ranges from $620\,\mathrm{nm}$ to $800\,\mathrm{nm}$. 
The additional bright spot $\approx 1\,\mathrm{\upmu m}$ to the left of the NV does not originate from a single emitter, but most likely from a broken off part of the AFM tip (see main text).
}
\label{fig:fig3}
\end{figure*}

For functionalization, samples with multiple straight waveguides and dimensions as derived in the last section ($m=400\,\mathrm{nm}, h=800\,\mathrm{nm}, w=500\,\mathrm{nm}$ in the coupling area) were fabricated and characterized. A waveguide was then selected and functionalized with a preselected single NV center hosted in a nanometer sized diamond, as described in \autoref{sec:wgdesign}
. After AFM manipulation, the presence of the NV center is verified in an all-confocal configuration where the the continuous wave $532\,\mathrm{nm}$ pump laser is focused onto the sample using a $\text{NA}=0.9$ objective lens (Olympus, MPLAPON60X). The reflected light is spectrally filtered by two $620\,\mathrm{nm}$ longpass filters (Omega optics) and spatially by a pinhole. The remaining fluorescence is directed either onto two avalanche photo diodes (APDs; Perkin-Elmer) in a Hanbury Brown and Twiss (HBT) configuration to verify the single photon generation, a CCD camera or a spectrometer. In the confocal scan, the NV center can be identified as a bright fluorescent spot near the waveguide (see upper part of \hyperref[fig:fig3]{\autoref{fig:fig3}(a)}). The lower part of \hyperref[fig:fig3]{\autoref{fig:fig3}(a)} shows the resulting second order correlation, measured with the HBT setup. Fitting the autocorrelation data to a three-level model:
\begin{equation}
g^{(2)}(\tau)=1 - (1+A)e^{-\frac{\tau}{t_1}}+Ae^{-\frac{\tau}{t_2}}
\end{equation}
where $A$ denotes the bunching amplitude, $t_1$ the antibunching time and $t_2$ the bunching time, we obtained a value of $g^{(2)}(0)=0.19\pm 0.01<0.5$ for this confocal excitation and detection scheme, which was limited by the APDs dark counts, its timing resolutions, and possible fluorescent background caused by residual graphite on the nanocrystal's surface.

The bright spot approximately $1\,\mathrm{\upmu m}$ left of the NV center does not originate from NV fluorescence as determined from the photon statistics, it most likely is some broken off part of the AFM tip used to position the NV center. Fortunately, this scatterer does not have a negative influence on the performance of the device.

To verify the coupling of the waveguide mode to the NV center, we couple the green pump laser to one input port of the waveguide via a lensed fiber and use the confocal microscope only to detect the resulting fluorescence. By scanning the confocal detection over the sample, the fluorescence map in \hyperref[fig:fig3]{\autoref{fig:fig3}(b)} is recorded, which shows that no background fluorescence of the waveguide can be detected even with strong pumping ($P_{ex} \approx 20\,\mathrm{mW}$ after the lensed fiber). The autocorrelation data originating from the bright spot reveals a $g^{(2)}(0)=0.18\pm 0.02<0.5$, which is comparable to the $g^{(2)}(0)$ obtained during confocal excitation of the NV center, verifying that the NV center can also efficiently be excited via the waveguide without increasing background fluorescence, thanks to the excellent optical properties of the SiO$_2$.

In a third configuration, the NV center is optically pumped via the microscope objective and the NV fluorescence directly coupled to the waveguide is detected by off-chip coupling the light from both waveguide ends to single-mode fibers. The remaining pump light is filtered out by a single $620\,\mathrm{nm}$ longpass filter at the end of each fiber before being detected by two APDs. \hyperref[fig:fig3]{\autoref{fig:fig3}(c)} shows the fluorescence map detected 
from one output arm of the waveguide
when scanning the laser over the waveguide and NV. In this configuration the waveguide itself acts as an intrinsic beamsplitter. Here the autocorrelation function recorded between the two output arms shows an anti-bunching with $g^{(2)}(0)=0.23\pm 0.03<0.5$ clearly indicating the mainly single-photon characteristics of the guided light in the waveguide.

Next, we assess the experimentally achieved dipole-waveguide coupling efficiency through the comparison of the fluorescence counts $F$ detected in the confocal and waveguide configurations, each corrected for its respective coupling efficiency $\beta$ and transmission efficiency $T$:
\begin{equation}\label{eq:coupling}
\frac{F_\mathrm{wg}}{T_\mathrm{wg} \cdot \beta_\mathrm{wg}} = \frac{F_\mathrm{conf}}{T_\mathrm{conf} \cdot \beta_\mathrm{conf}}.
\end{equation}

Our confocal setup has an overall collection efficiency of $T_\mathrm{conf} \cdot\beta_\mathrm{conf} =8.4\pm 1.7 \,\%$ (fraction of dipole emission collected by the objective lens $\beta_\mathrm{conf} =12\pm 2 \,\%$, transmittance of optical elements $T_\mathrm{conf}= 67\pm 1\,\%$) and provides typical count rates of $F_\mathrm{conf}=50\pm 5\,\mathrm{Kcts/s}$ from single NV centers \cite{Pyrlik2019}. The integrated waveguide system has a transmission of $T_\mathrm{wg}= 35\pm 1\,\%$ and provides total count rates of $F_\mathrm{wg}=5.0\pm 0.5\,\mathrm{Kcts/s}$ per output arm. Given that the total counts emitted by the NV stays the same in both situations, rearranging \autoref{eq:coupling} results in a NV-waveguide coupling efficiency of
\begin{equation}
\beta_\mathrm{wg} = \frac{F_\mathrm{wg}}{T_\mathrm{wg}} \cdot \frac{T_\mathrm{conf} \cdot \beta_\mathrm{conf}}{F_\mathrm{conf}} = 4.8\pm 1.2 \,\%.
\end{equation}

This derived coupling efficiency fits very well to the simulated values of $\sim 6\,\%$ ($\sim 4\,\%$) for a horizontally (vertically) oriented dipole emitter coupling to the TE (TM) mode in the waveguide.

\section{\label{sec:increasing}Increasing coupling efficiency}

\begin{figure}[htp]
  \centering
  \includegraphics[width=0.75\columnwidth]{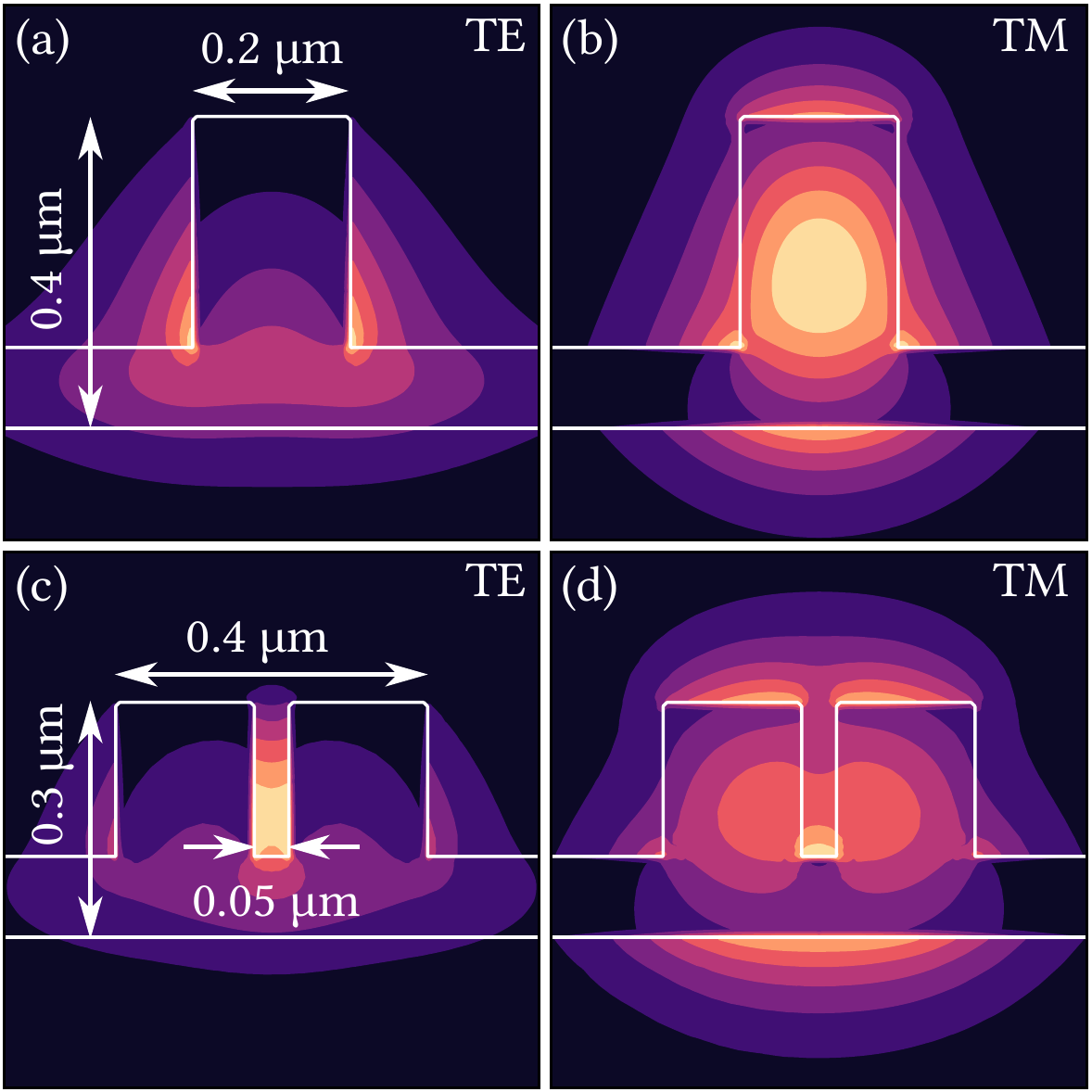}
\caption{
\textit{Increasing coupling efficiency.}
In order to increase coupling efficiency of the NV emitter to the guided mode in the waveguide, a structure with a thinner supporting membrane ($m=100\,\mathrm{nm}$) is proposed, enabling better access to the evanescent field. 
The field profile ($|E|^2$) of the guided TE and TM mode for two different optimized designs is shown. With the traditional rib design in (a) and (b) the $\beta$ factor can be increased to $15\,\%$ and $10\,\%$. In (c) and (d) a slot of width $w_s=50\,\mathrm{nm}$ is added and the $\beta$ factor can be increased to $65\,\%$ and $32\,\%$ for the TE and TM mode coupling to a vertical and horizontal emitter, respectively.}
\label{fig:fig5}
\end{figure}

For demanding applications, e.g. the experimental realization of a single photon non-linear device \cite{Kewes2016}, an increased coupling efficiency between emitter and guided mode would be desirable. In the following we will comment on some prospects for increasing the coupling efficiency far beyond the theoretical value of $\sim 5\,\%$  of our fabricated device, but at the same time have higher manufacturing and/or assembly requirements.

In principle, it is crucial to obtain better access to the guided mode in the waveguide, i.e. smaller waveguide dimensions, in order to make the evanescent field more accessible.
In \autoref{sec:guidedmodes} we have already optimized the dimensions of the waveguide, but with a very conservative thickness of the supporting membrane ($m=400\,\mathrm{nm}$) to ensure a very robust device. If, for example, the membrane thickness is reduced to $m=100\,\mathrm{nm}$, parameter scans yield the optimal dimensions of the waveguide ($w=200\,\mathrm{nm}$ and $h=400\,\mathrm{nm}$, see \hyperref[fig:fig5]{\autoref{fig:fig5}(a,b)}) which allows an increased $\beta$ factor of $15\,\%$ ($10\,\%$) for a vertical (horizontal) dipole emitter coupling to the guided TE (TM) mode.

In order to improve the coupling efficiency even further, a slot-waveguide could be implemented. Here, the supported mode channels most of the electromagnetic field within the slot, allowing optimal access to the guided mode when the emitter is placed in the slot. In \hyperref[fig:fig5]{\autoref{fig:fig5}(c,d)} we show an optimized design for a slot-waveguide with a slot width of $w_s=50\,\mathrm{nm}$ and a membrane thickness of $m=100\,\mathrm{nm}$. This design is more difficult to fabricate and functionalize, but would also result in an greatly increased $\beta$ factor of $65\,\%$ ($32\,\%$) for a vertical (horizontal) dipole emitter located in the center of the slot and coupling to the TE (TM) guided mode.

A further increase of the coupling efficiency could be achieved e.g. by using an even thinner membrane or an even smaller slot width. It should be noted, that this requires even higher fabrication, assembly and handling demands, as the device becomes more fragile. Another prospect for increasing the coupling efficiency could be more complex systems, e.g. overgrowing the emitter with SiO$_2$ after positioning it or an adiabatic coupling of the dipole emitter to the waveguide \cite{Patel2016,Davanco2017}.

\section{\label{sec:conclusion}Conclusion and outlook}
In summary, we present on-chip SiO$_2$ photonic structures with ultra-low fluorescence at visible wavelengths which are very well suited for the integration of solid state single-photon emitters that require relatively high excitation powers, such as the NV center. The integrated structure allows effective routing of excitation laser sources and single photons via a freestanding rib-waveguide configuration. The on- and off-chip power coupling efficiency to single mode fibers can be increased by introducing 2D tapered sections of the waveguide to adiabatically transform the size of the guided mode. By deterministically positioning a nanodiamond with a single NV center in close proximity to the waveguide, we were able to demonstrate single photon generation by either pumping the NV center over the waveguide and detection in free-space, or by pumping over the microscope objective and detection of the photons coupled to the single guided mode of the waveguide. We could also experimentally verify the theoretically predicted NV-waveguide coupling efficiency.
Furthermore we presented and commented on possible ways to enhance the theoretical coupling efficiency between the guided mode and the quantum emitter by over one order of magnitude.

The device and functionalization presented here is not limited to NV centers in diamond, but can also be transferred to other solid-state quantum emitters in the visible, such as other defect centers in diamond or defects in 2D materials. 
Furthermore, the integrated device can be extended with other on-chip photonic structures such as high-Q ring resonators \cite{Pyrlik2019}, 
directional couplers, on-chip detectors \cite{Ferrari2018}, and microwave antennas which allows for the monolithic realization of complex devices consisting of various functionalities, e.g. the optical microintegration of those chips with high functional density together with light sources, detectors, and electronics.

In this way, the presented photonic platform and integration technique opens up the possibility of building up complex structures with several integrated functionalities.
Even the packaging of the chip and pump laser diodes within an optical module can be envisioned. This would be highly attractive for compact modules for integrated quantum technologies.

\section{Acknowledgments}
We thank G\"unter Kewes and Bernd Sontheimer for fruitful discussions.
We acknowledge financial support by the European Fund for Regional Development of the European Union in the framework of project iMiLQ, administrated by the Investitionsbank Berlin within the Program to Promote Research, Innovation, and Technologies (ProFIT) under grant 10159465 and the Federal Ministry of Education and Research of Germany in the framework of Q.Link.X (project number 16KIS0876).
\bibliography{Mendeley.bib}

\end{document}